\documentclass[12pt]{article}
\usepackage{multirow}
\usepackage{lipsum} 
\usepackage{booktabs} 
\usepackage{wrapfig} 
\usepackage{graphicx}
\usepackage{amsmath}
\usepackage{amsfonts}
\usepackage{amssymb}
\usepackage{natbib}
\usepackage{soul}
\usepackage{graphicx}
\usepackage{graphicx}

\usepackage[dvipsnames,usenames]{color}

\usepackage{amsmath}                        

\definecolor{DarkGreen}{rgb}{0.5,0.8,0.6}   
\definecolor{RGBblack}{rgb}{0.0,0.0,0.0}    

\definecolor{grau}{rgb}{0.8,0.8,0.8}

\newcommand{\chen}[1]{\color{orange}}

\usepackage{algorithm}
\usepackage{algorithmic}
\usepackage{mathtools}
\usepackage{multirow}
\usepackage{bbm}
\usepackage{enumitem}
\usepackage{tikz}
\usepackage[title]{appendix}
\usepackage{subcaption}
\usepackage[flushleft]{threeparttable}
\usepackage{color,bm,xcolor}

\DeclareMathOperator*{\argmax}{arg\,max}
\DeclareMathOperator*{\argmin}{arg\,min}

\title{PoD-BIN:  A Probability of Decision Bayesian Interval Design for Time-to-Event  Dose-Finding Trials with Multiple Toxicity Grades  }
\author{Meizi Liu \footnote{Department of Public Health Sciences, The University of Chicago, Chicago, USA}, Yuan Ji \footnote{Department of Public Health Sciences, The University of Chicago, Chicago, USA}, Ji Lin \footnote{Sanofi US}}
\date{}

\begin{document}

\maketitle

\section*{Abstract}
We consider a Bayesian framework based on ``probability of decision'' for dose-finding trial designs. The proposed PoD-BIN design evaluates the posterior predictive probabilities of up-and-down decisions. In PoD-BIN, multiple grades of toxicity, categorized as the mild toxicity (MT) and dose-limiting toxicity (DLT), are modeled simultaneously, and the primary outcome of interests is time-to-toxicity for both MT and DLT. This allows the possibility of enrolling new patients when previously enrolled patients are still being followed for toxicity, thus potentially shortening trial length. The Bayesian decision rules in PoD-BIN utilize the probability of decisions to balance the need to speed up the trial and the risk of exposing patients to overly toxic doses. We demonstrate via numerical examples the resulting balance of speed and safety of PoD-BIN and compare to existing designs. 


\section{Introduction}\label{intro}
\subsection{Overview}
Phase I clinical trials are the first-in-human studies evaluating the safety and tolerability of a new treatment. Given a series of ascending dose levels, the goal of phase I clinical trial is to identify a maximum tolerated dose (MTD) defined as the highest dose having a dose-limiting toxicity (DLT) probability close to or lower than a target toxicity rate, say 0.3. Under this goal, previous designs for phase I trials can generally be categorized into two classes: $(i)$ algorithm-based designs, such as the 3+3 \citep{storer1989} design and the i3+3 design \citep{Liu2020}, and $(ii)$ model-based designs, such as the continual reassessment method (CRM) \citep{o1990continual}, the Bayesian logistic regression method (BLRM) \citep{neuenschwander2008critical}, and the interval-based designs like mTPI \citep{ji2010modified}, mTPI-2 \citep{guo2017bayesian}, and keyboard \citep{yan2017keyboard}. 

In conventional phase I trials, patients are enrolled sequentially, and doses are assigned for successive patients based on outcomes observed within a certain assessment window. With recent advances in targeted therapy and immunotherapy in oncology, the problem of late-onset toxicity has been a major concern in phase I trials as the toxicities associated with these new treatments can take longer to observe than traditional cytotoxic therapies \citep{kanjanapan2019delayed}. In addition, clinically it is important to distinguish different grades of toxicities instead of a binary DLT outcome, usually referring to grade 3 or higher grade toxicities \citep{CTCAE}.   For example, in the phase I trial that tests the toxicity of certinib \citep{cho2017ascend}, grades 1 and 2 adverse events (AEs) at the 750mg dose have led to concomitant medication, study drug intervention, or dose reduction. These AEs would not be categorized as a DLT, although they still lead to a dose reduction to 450mg in the trial, which shows comparable PK exposure and a more favorable safety profile. It is, therefore, desirable to consider the safety of a dose based on its toxicity beyond the conventional definition of DLT -- defined by dichotomizing severity of toxicities into a binary variable with a significant loss of toxicity information obtained from each patient. 

\subsection{Existing approaches on time-to-toxicity and multi-grade toxicity models}
We attempt to shorten the trial duration and better characterize the toxicity profile of a dose by modeling 
time-to-event measurements for toxicity outcomes of multiple toxicity grades. To shorten the duration of phase I trials, several designs have proposed rolling enrollment of future patients 
while some existing patients in the trial are still pending for DLT assessment. \cite{cheung2000sequential} introduce the time-to-event CRM (TITE-CRM) design which considers a weighted likelihood approach, where a patient with complete outcomes receives full weight, and a patient with pending outcome receives a partial weight based on the time of follow-up. Both the expectation maximization CRM (EM-CRM) \citep{yuan2011robust} and data augmentation CRM (DA-CRM) \citep{liu2013bayesian} adopt the idea of missing data modeling, where the missing/pending outcomes are explicitly considered.  \cite{yuan2018time} propose the time-to-event BOIN (TITE-BOIN) design to accommodate late-onset toxicity by imputing the unobserved pending toxicity outcome using the single mean imputation method. The rolling TPI (R-TPI) design is an extension of the mTPI-2 design with a fixed set of decision rules to accommodate pending outcomes \citep{guo2019r}. Recently, in the probability-of-decision toxicity probability interval design (PoD-TPI) by \cite{zhou2020pod}, the dose assignment decision is treated as a random variable, and the posterior distribution of the decisions is computed to accommodate uncertainties in pending outcomes.

Separately, several methods have been proposed to model different toxicity grades in the estimation of MTD based on non-time-to-event observations. \cite{bekele2004dose} first introduce a concept of total toxicity burden (TTB), which is a weighted sum of the toxicity grades across different toxicity types. \cite{yuan2007continual} propose the quasi-CRM method, in which toxicity grades are first converted to the equivalent toxicity score, and then incorporated into the CRM using the quasi-Bernoulli likelihood to find MTD. \cite{lee2010continual} propose the continual reassessment method with multiple constraints (CRM-MC), which allows for the specification of various toxicity thresholds. \cite{iasonos2011incorporating} introduce a two-stage CRM that incorporates lower-grade toxicities in the first stage and separately models the rates of DLT and lower-grade toxicities in the second stage. \cite{van2012dose} propose the proportional odds CRM, incorporating toxicity severity with proportional odds models. Later, \cite{ezzalfani2013dose} define the total toxicity profile (TTP) as the Euclidean norm of the weights of toxicities experienced by a patient and characterize the overall severity of multiple types and grades of toxicities.

Recently, \cite{Lee2019} develop the TITE-CRMMC (Time to Event Continual Reassessment Method with Multiple Constraints) design that incorporates time-to-event information with multiple toxicity grades, combining the TITE-CRM \citep{cheung2000sequential} and CRM-MC \citep{lee2010continual} designs. It takes into account partial information on both moderate toxicities (MTs) and DLTs observed during a trial. However, TITE-CRMMC assumes that once a patient experiences a DLT, having additional moderate toxicity would no longer be counted as a toxicity outcome. In other words, patients with DLT  only  and patients with both moderate toxicity and DLT are not differentiated.

In this paper, we propose a Bayesian design, namely the  probability-of-decision Bayesian interval design (PoD-BIN), to accelerate phase I trials by modeling time to different toxicity grades.  Specifically, time-to-toxicity for both DLT and MT are modeled using a latent probit regression, and a summary statistic called the ``toxicity burden" is used to summarize the total toxicity impact to a patient. Importantly, dose-finding decisions are treated as a random variable and statistical inference is based on the posterior predictive distribution of pending toxicity outcomes. Innovatively, we compute and threshold probability of decisions to control the risk of making aggressive decisions. The remainder of this paper is organized as follows. In Sections \ref{met} and \ref{des}, we describe the probability model and propose the PoD-BIN design, respectively. In  Section \ref{sim},  we present simulation studies and results. The paper ends with a discussion in Section \ref{dis}.

\vspace{-0.5cm}
\section{Probability model and Inference}\label{met}
\subsection{Probability model}\label{prob}
Consider estimating the probabilities of DLT and MT among a set of $D$ dose levels, $d \in \{1,2,\dots, D\}$. Based on the Common Terminology Criteria for Adverse Events (CTCAE) by \cite{CTCAE}, general guidelines for all possible adverse events describe grade 1 as ``mild,'' 2 as ``moderate,'' 3 as ``severe,'' 4 as ``life-threatening or disabling,'' and 5 as ``death.'' We define MT as mild or moderate toxicities that are usually not considered as DLTs (i.e., grade $1-2$ based on CTCAE criteria), and DLT as the pre-specified severe adverse events usually with grade $3-5$ by CTCAE. In practice, individual trials may extend CTCAE to define their own MTs and DLTs. 

Let $W$ denote the length of the pre-specified assessment window. In oncology, $W$ is usually set to 21 to 28 days, corresponding to the length of one treatment cycle. 
Now assume a sample size of $N$ patients is to be enrolled in a trial. We denote $v_i$ the follow-up time for patient $i$, for $i=1, \dotsc ,N$, and $x_i$ the dose assigned to patient $i$. Let $T_{MT,i}$ be the time to the manifestation of the first MT event for patient $i$, and $T_{DLT,i}$ the time to the manifestation of the first DLT event.  We introduce $Y_i$ as an ordinal variable with four levels, summarizing the MT and DLT outcomes patient $i$. Suppose patient $i$ has been followed for time $v_i$, Define an outcome variable as 
\begin{equation}
 Y_i^{(v_i)} = 
    \left\{  \begin{array}{lr}
        1, & \text{if \ } T_{MT,i}>v_i,   T_{DLT,i}>v_i,\\
        2, & \text{if \ } T_{MT,i} \leq v_i,   T_{DLT,i}>v_i,\\
        3, & \text{if \ } T_{MT,i}>v_i,   T_{DLT,i}\leq v_i,\\
        4, & \text{if \ } T_{MT,i}\leq v_i,   T_{DLT,i}\leq v_i.
        \end{array} \right.
\end{equation}
In words,  $Y_i$ takes a value of $1$ if no toxicity is observed by time $v_i$, $2$ if only MT is observed, $3$ if only DLT is observed, and $4$ if both MT and DLT are observed by time $v_i$. By definition, $Y_i$ is ordinal and the higher its value the more severe the toxicity. Also, $Y_i$ is a function of follow-up time $v_i$, and therefore it is time-varying as patient $i$ continues to be followed. When patient $i$ completes follow-up, define $Y_i^*$ as the ordinal outcome at the end of assessment window, i.e.,
\begin{equation}\label{eq0}
 Y^*_i = 
    \left\{  \begin{array}{lr}
        1, & \text{if \ } T_{MT,i}>W,   T_{DLT,i}>W,\\
        2, & \text{if \ } T_{MT,i} \leq W,   T_{DLT,i}>W,\\
        3, & \text{if \ } T_{MT,i}>W,   T_{DLT,i}\leq W,\\
        4, & \text{if \ } T_{MT,i}\leq W,   T_{DLT,i}\leq W,
        \end{array} \right.
\end{equation}
where $W$ is the duration of the assessment window. Since $W$ is a fixed value, such as 21 days, $Y_i^*$ is not time-varying. 

In a traditional phase I trial, DLT$=\mathbbm{1}\{Y^*_i=3 \text{ or } 4\}$ and no DLT$=\mathbbm{1}\{Y^*_i=1 \text{ or } 2\}$. We first model the relationship between the dose level $x_i$ and the time-to-toxicity $Y^*_i,$  and then the relationship between $Y^{(v_i)}_i$ and $Y^*_i$. We assume a latent Gaussian variable $Z$ that underlies the generation of the ordinal response $Y_i^*$, given by a regression of $Z_i$ on dose $x_i$ as
\begin{equation}\label{eq:probit}
   Z_i=x_i\beta+\epsilon,
\end{equation}
where $\epsilon$ is the error term drawn from a standard Gaussian distribution,  i.e.,  $Z_i\sim N(x_i\beta, 1)$.  We assume the effect of doses, $\beta \in \mathbb{R}^+$, is positive across the four ordered toxicity categories to enforce monotone toxicity over doses. Therefore, let $\Phi(\cdot)$ denote the distribution function of a standard Gaussian random variable, and by Equation~(\ref{eq0}) we define the probability of toxicity outcome $Y_i^*=c$ at dose $d$ as  
\begin{equation}\label{pdj}
 p_{d,c} \triangleq Pr(Y_i^*=c|x_i=d) = Pr(g_{c-1} \leq Z_i < g_c) = \Phi (g_c - d\beta) - \Phi (g_{c-1} - d\beta), 
\end{equation}
where  $c= 1,\dotsc,4$, the parameter vector $\boldsymbol{g} = (g_0, g_1, \dotsc, g_4)$ are the cutoff values, and $-\infty = g_0 < g_1 < \dotsb < g_4 = + \infty$. We also require $g_1\equiv 0$ to ensure identifiability \citep{johnson2006ordinal}. 

Suppose a total of $n$ patients have been enrolled, and denote $\{Y_i^{(v_i)}\}$ as the currently observed data. The likelihood function can be written as
 \begin{equation}\label{eq1} 
 L_n(\beta,\boldsymbol{g},\boldsymbol{Z} | \boldsymbol{y, v, x})=\prod_{i=1}^n \prod_{ k=1}^{4} [Pr(Y_i^{(v_i)}=k|x_i)]^{\mathbbm{1}(Y_i^{(v_i)}=k)}.
 \end{equation}
To proceed with Bayesian modeling,  we express $Pr(Y_i^{(v_i)}=k \mid x_i)$ in  
Equation~(\ref{eq1}) as a function of model parameters, $\boldsymbol{g}$ and $\beta$, using the latent probit regression in Equation~(\ref{pdj}). Specifically, write
\begin{equation}\label{eq2}
\begin{aligned}
Pr(Y_i^{(v_i)}=  k  | x_i=d) 
                       &= \sum_{c=1}^4Pr(Y_i^{(v_i)}=  k  |Y_i^{*}=  c  )Pr(Y_i^{*}= c|x_i =d) \\
   & = \sum_{c=1}^4 w_{k, c} \cdot p_{d, c}   \\
&= \sum_{c=1}^4 w_{k, c} \cdot \{\Phi (g_c - d \beta) - \Phi (g_{c-1} - d \beta) \},
\end{aligned}
\end{equation}
where $w_{k, c}=Pr(Y_i^{(v_i)}=k |Y_i^{*}=c)$  is a conditional probability and usually modeled as a weight function in time-to-event models \citep{cheung2000sequential, Lee2019}. Denote $\boldsymbol{W}=[w_{k,c}; k,c=1,\dots,4]$ as a $4\times 4$ matrix given by
\[
\boldsymbol{W}=\begin{bmatrix}
    w_{1,1} & \dots & w_{1,4} \\
    \vdots  & \ddots & \vdots \\
    w_{4,1} & \dots & w_{4,4}
  \end{bmatrix}.
\]
Therefore, Equation~(\ref{eq2}) can be rewritten as

\begin{align}\label{eq3}
\begin{bmatrix}
   Pr(Y_i^{(v_i)}=1|x_i)  \\
    \vdots   \\
    Pr(Y_i^{(v_i)}=4|x_i)
  \end{bmatrix} = \boldsymbol{W} 
   \begin{bmatrix}
    Pr(Y_i^{*}=1|x_i) \\
    \vdots   \\
    Pr(Y_i^{*}=4|x_i) 
  \end{bmatrix}. 
\end{align}

In $\boldsymbol{W}$, $w_{k, c}=0$ if $k> c$ since it is impossible to observe a toxicity outcome by an early time point but not later. By the same argument, $w_{1,1}=1$. Following the recommendation in TITE-CRM \citep{cheung2000sequential}, the linear weight function yielded similar operating characteristics compared to other complicated weight functions, and therefore we assume the $w_{k,c}$'s are linear in time, i.e., $Pr(T_{MT,i}\leq v_i|T_{MT,i}\leq W)=  \frac{v_i}{W}$. 
We derive the entire $\boldsymbol{W}$ matrix based on the linearity assumption given by 
\begin{align}\label{eq3}
\boldsymbol{W}=
  \begin{bmatrix}
    1 & 1-\frac{v_i}{W} & 1-\frac{v_i}{W} & (1-\frac{v_i}{W})^2 \\
    0 & \frac{v_i}{W} & 0 & \frac{v_i}{W}(1-\frac{v_i}{W}) \\
    0 & 0 & \frac{v_i}{W} & \frac{v_i}{W}(1-\frac{v_i}{W}) \\
    0 & 0 & 0 & \frac{v_i^2}{W^2}
  \end{bmatrix}.
\end{align}
See details in Web Appendix A. 

\subsection{Inference}\label{inference}

Treating $\boldsymbol{Z}$ as augmented data, posterior inference is conducted by sampling 
the model parameters $(\beta, \boldsymbol{g}, \boldsymbol{ Z})$. We fix the first cutoff parameter $g_1$ at $0$, and assume a flat prior $f(\boldsymbol{g})\propto 1$ on $\boldsymbol{g}=( g_2, g_3)$ and a truncated flat prior $f(\beta) \propto \mathbbm{1}(\beta>0)$ on $\beta$ over the positive real  values. With the likelihood function in Equation~(\ref{eq1}) and the prior models, we conduct posterior inference on the parameters $(\beta, \boldsymbol{g})$ and the latent variable $\boldsymbol{Z}$ using the Metropolis-Within-Gibbs algorithm \citep{johnson2006ordinal}, a technique of \mbox{Markov-chain} Monte Carlo (MCMC) simulation (Web Appendix B). The ordering constraints on the cutoffs  $\boldsymbol{g}=(g_2, g_3)$ are imposed through the sampling step in MCMC.

One drawback of the augmented data approach for ordinal data is that MCMC has poor numeric stability, especially when the numbers observed at each ordinal category are small. As such, we implement a two-stage design -- in the first stage, only complete follow-up data will be used until sufficient information from patients has been acquired. This means no time-to-event modeling, and we will introduce a special algorithm for the first stage. The second stage allows for rolling enrollment and dose assignments while toxicity outcomes of enrolled patients are still pending. The dose assignment decisions are made based on the posterior predictive distribution of the toxicity outcomes for the pending patients.  The two-stage design will become clear later on.

\subsection{Posterior Predictive Imputation}\label{imputation}

We follow the idea of the PoD-TPI design \citep{zhou2020pod} for decision-making. Specifically,  the observed time-to-event data  
can be used to predict the probability of a pending patient experiencing MT or DLT 
at the end of follow-up,  under which we can compute the probability of possible dose allocation decisions. These probabilities of decisions (PoDs) are  used to guide dose allocation decisions. We discuss details next. 

At a given moment of the trial, suppose a total of $n$ patients have been  enrolled,  and the current dose used for treating patients in the trial is $d$.  Recall that $Y^{(v_i)}_i$ denotes the observed toxicity outcome by follow-up time $v_i$ and $x_i$ denotes  the assigned dose level for   patient $i$. Let 
$M_i = 1$ if patient $i$ is still being followed without definitive toxicity outcome, i.e., $v_i < W$; and let $M_i = 0$  if the patient finishes the follow-up, i.e., $v_i=W$. In missing data literature, $M_i$ is the missingness indicator.  
Denote $\mathcal{I}_d = \{i : x_i = d, M_i = 1\}$ the index set of the pending patients treated at dose $d$.  Using this notation, we define the outcomes of the patients as follows.  At dose $d$, let $S_d = \{Y^*_i : i \in \mathcal{I}_d\}$  be the set of future unobserved outcomes of pending patients when they complete the follow-up. And let $C_d = \{Y^*_i : i \notin \mathcal{I}_d\}$ be the set of observed outcomes of patients who have completed follow-up. If $S_d$ were known, then given a statistical design, the dose-finding decision would be given by $A_d = \mathcal{A}(S_d,C_d),$ where $\mathcal{A}$ represents a statistical design using complete data $(S_d,C_d)$. However, since $S_d$ is not observed, it is treated as random, which in turn makes $A_d$ a random variable. With proper probability modeling of $S_d$, we can then calculate the probability of $A_d$, which is the PoD.  

Recall $p_{d,c} = Pr(Y_i^{*}= c | x_i =d)$  is the marginal probability of experiencing toxicity outcome $c\in\{1,\dots,4\}$ at dose level $d$ when a patient completes  the follow-up.   
 Denote  $\boldsymbol{p_d}=(p_{d,1},p_{d,2},p_{d,3},p_{d,4})$. The proposed model relies on the estimation of $\boldsymbol{p_d}$ using the observed data $\{Y_i^{(v_i)}, v_i\}$, which can be connected by the following predictive probability. In particular, for patient $i$ who has been followed for time $v_i$ with outcome $\{Y_i^{(v_i)}\}$, the predictive probability for future outcome $Y_i^*$ at the end of follow-up is given by  
 \begin{align}\label{eq4}
       Pr(Y_i^*=c | Y_i^{(v_i)}=k, x_i=d)
         = &\int \underbrace{ Pr(Y_i^*=c | Y_i^{(v_i)}=k,x_i=d, \boldsymbol{p_d})}_{I} \pi(\boldsymbol{p_d} |  \text{data}  ) 
             d\boldsymbol{p_d}.
\end{align}

\noindent where $\pi(\boldsymbol{p_d} |  \text{data}  )$ 
is the posterior distribution of $\boldsymbol{p_d}$. For $I$, using Bayes' theorem, we express it as a function of $\boldsymbol{p_d} $ 
as follows:
 \begin{align*}
         I 
        =& \frac{Pr(Y_i^{(v_i)}= k |Y^{*}_i=c) Pr(Y^{*}_i=c|x_i=d,\boldsymbol{p_d})}{\sum_{h=1}^4Pr(Y_i^{(v_i)}= k |  Y^{*}_i=h)Pr(Y^{*}_i=h|x_i=d, \boldsymbol{p_d}) }\\
      =&\frac{w_{k, c} p_{d,k}}{\sum_{h=1}^4 w_{k, h} p_{d,h}}. 
\end{align*} 
Therefore, Equation (\ref{eq4}) can be estimated by posterior inference using the latent probit model. Specifically, we obtain MCMC samples of $\boldsymbol{g}$ and $\beta$, denoted as $\{\boldsymbol{g}^{(b)}, \beta^{(b)}, b=1,\dots,B\}$. 
Therefore, the posterior predictive probability in Equation (\ref{eq4})  can be approximated as, 
\begin{equation}
         Pr(Y_i^{*}=c|  Y_i^{(v_i)}= k, x_i=d)
        = \frac{1}{B} \sum_{b=1}^B \frac{ w_{k, c} \hat{p}_{d,k}^{(b)}}{\sum_{h=1}^4 w_{k, h} \hat{p}^{(b)}_{d,h}},
\end{equation}
where $\hat{p}^{(b)}_{d,c} = \Phi(\hat{q}_c^{(b)} - d\hat{\beta}^{(b)}) - \Phi(\hat{q}_{c-1}^{(b)} - d\hat{\beta}^{(b)}).$ 
 
Given the predictive probability of the $i$th patient, we can easily derive the probability of observing $S_d = s$ at dose $d$, where $s \in \{(s_1,\dots,s_J); s_j \in \{1,2,3,4\} \}$ is the set of possible future outcomes for the pending patients and $J_d=|\mathcal{I}_d|$ is the number of pending patients at dose $d$. Assuming independence across patients' outcomes, we have
\begin{equation}
Pr(S_d=s|Y_i^{(v_i)},x_i) = \prod_{i\in \mathcal{I}_d} Pr(Y^*_i = s_j|Y_i^{(v_i)},x_i) , \ j=1,\dots,J_d. 
\end{equation}
\noindent Finally, we define the PoD as 
\begin{equation}\label{eq9}
Pr(A_d=a|Y_i^{(v_i)},x_i) = \sum_{s:\mathcal{A}(s,C_d)=a}Pr(S_d=s|Y_i^{(v_i)},x_i), 
\end{equation}
where decision $a$ 
denotes the dose-finding decision based on a dose-finding algorithm $\mathcal{A}$ for complete data. An algorithm $\mathcal{A}$ is introduced in Section \ref{algorithm} next. 
Based on PoDs, the proposed PoD-BIN design assigns patients to a dose according to the decision $A^*_d$ with the largest PoD, i.e.,
\begin{equation}\label{eq10}A^*_d = \argmax_{a} Pr(A_d=a|Y_i^{(v_i)},x_i).\end{equation}

\vspace{-0.5cm}
\section{Trial Design}\label{des}
\subsection{ Target toxicity }

\noindent The goal of the proposed PoD-BIN design is to incorporate the time-to-event of both MT and DLT outcomes into the dose-escalation decisions. We first combine the MT and DLT outcomes in the form of a weighted average, representing the toxicity burden of each patient and dose. The concept of toxicity burden was first introduced by \cite{bekele2004dose}.  Toxicity burden 
requires elicitation of positive-valued numerical weights with a higher weight corresponding to greater severity, where the weights characterize the relative extent of harm that is associated with experiencing the toxicity at the given severity level in relation to the other severity levels. These weights are elicited based on physicians' consensus at the design stage of a trial. In this case, we denote $r_M$ as the \textit{relative severity weight} for MT relative to DLT. For example,  $r_M=0.2$  is  
interpreted as  that  the impact of experiencing five MT events on patients' health is equivalent to that of experiencing one DLT event. We define the toxicity burden (TB) at dose $d$ as 
\begin{equation}
TB_d= r_M{p}_{MT,d} + {p}_{DLT,d},
\end{equation}
 where ${p}_{MT,d},{p}_{DLT,d}$ are the probabilities of MT and DLT at dose $d$,  defined as 
\begin{equation}
\begin{aligned}
 p_{MT, d} &= Pr(Y^*_i = 2 |x_i=d) + Pr(Y^*_i = 4 |x_i=d)=p_{d,2}+p_{d,4} , \ \text{and} \\  p_{DLT, d} &= Pr(Y^*_i = 3 |x_i=d)  + Pr(Y^*_i = 4 |x_i=d)=p_{d,3}+p_{d,4},
 \end{aligned}
\end{equation}
 respectively. Let $p^*_{DLT}$ be the target DLT probability as in conventional phase I trials, for example, $p^*_{DLT}=0.3$ .  We further define the target toxicity burden as 
 \begin{equation}
 TTB=p^*_{DLT} + r_Mp^*_{MT},
 \end{equation}
 where $p^*_{MT}$ is the target MT probability. For simplicity, we set $p^*_{MT}=0$ in subsequent numerical examples, although it needs not to be $0$. Similar to $p^*_{DLT}$, $p^*_{MT}$ should be elicited with trial physicians based on the specific context of a trial.

\subsection{Dose assignment algorithms}\label{algorithm}
The proposed PoD-BIN design consists of two stages. The first stage is a simple interval-based design that ends until sufficient information from patients has been collected. Stage I requires that enrolled patients are fully followed with no pending outcomes, i.e., no rolling enrollment. At the second stage, the Bayesian time-to-event probability model introduced in Section \ref{prob} is applied, allowing new patients to be enrolled when previously enrolled patients have pending outcomes. In addition, the posterior predictive imputation of pending outcomes is performed, and patients are allocated to available doses according to the optimal PoD. 

\paragraph{Stage I }
 The proposed Stage I algorithm follows and extends the idea in the i3+3 design \citep{Liu2020} (Web Appendix C). Specifically, consider decisions based on an estimated toxicity burden $\widehat{TB_d}$ and an equivalence interval ($EI$) of the $TTB$. 
We define the $EI$ of $TTB$ as $(TTB-\epsilon_1, TTB+\epsilon_2)$ \noindent where $\epsilon_1, \epsilon_2$ are small constants that reflect investigators' tolerance of the deviation from $TTB$ for the MTD. In other words, $(TTB-\epsilon_1)$ is the lowest toxicity burden for which a dose can be considered as the MTD, and $(TTB+\epsilon_2)$ is the highest burden.  Suppose dose $d$ is the current dose used to treat patients. Consider an estimated toxicity burden, $ \widehat{TB}_d= r_M\frac{n_{MT,d}}{n_d} + \frac{n_{DLT,d}}{n_d},$
where $n_{d}$ is the total number of patients, and $n_{MT,d}\leq n_d$ and $n_{DLT,d}\leq n_d$ are the numbers of patients who experience MT and DLT at dose $d$, respectively. Define $\widehat{TB}_{d, -1}$ the value of $\widehat{TB_d}$ by assuming that the patient with the least severe toxicity outcome enrolled at the current dose $d$ had no toxicity at all. Mathematically, $\widehat{TB}_{d,-1}= r_M\frac{n_{MT,d,-1}}{n_d} + \frac{n_{DLT,d,-1}}{n_d},$ where $n_{MT,d,-1}=n_{MT,d}-1$ if the least severe patient experiences MT,  $n_{MT,d,-1}=n_{MT,d}$ otherwise. Similarly, $n_{M=DLT,d,-1}=n_{DLT,d}-1$ if the least severe patient experiences DLT,  $n_{DLT,d,-1}=n_{DLT,d}$ otherwise. Table \ref{tab:stage1} provides an example for the calculation of  $\widehat{TB}_{d}$ and  $\widehat{TB}_{d, -1}$. In the example, the patient with the least toxicity treated at the current dose $d$ is the first patient with only MT. Then the MT is removed and replaced by no toxicity in order to calculate  
$\widehat{TB}_{d, -1}$. In the case when the patient with the least toxicity has no toxicity, then $\widehat{TB}_{-1,d} =\widehat{TB}_{d}$. 

\begin{table}
\caption{An example of $\widehat{TB}_d$ and $\widehat{TB}_{d, -1}$ for $n_d=3$ patients, with $n_{MT,d}=2$ MT events and $n_{DLT,d}=2$ DLT events. Assume $r_M=0.2,$ which means a DLT event is the same as $1/0.2=5$ MT events in terms of harm to health. Top panel: the observed data and the corresponding toxicity burden $\widehat{TB}_d$. Bottom panel: the data with one MT (for patient 1) deleted and the corresponding toxicity burden $\widehat{TB}_{d, -1}$.} \label{tab:stage1}
\begin{center}
\small
\begin{tabular}{ccccc}
\hline
\multirow{2}{*}{Patient $\#$}& \multicolumn{3}{c}{Observed data}    & \multirow{2}{*}{$\widehat{TB}_{d} =r_M\frac{n_{MT,d}}{n_d} + \frac{n_{DLT,d}}{n_d}$ }    \\
  & &  MT & DLT  & \\ \cline{1-5}
1& &  \bf{1} & 0 & \\   

2 &  &1 & 1 & $ 0.2 \times \frac{2}{3} + \frac{2}{3} = 0.8$ \\

3 & & 0 & 1  & \\
\hline 
$n_d=3$& & $n_{MT,d}= 2$ & $n_{DLT,d}=2$\\ \hline 
\end{tabular}  

\vspace{3em}

\begin{tabular}{ccccc} \hline
\multirow{2}{*}{Patient $\#$}& \multicolumn{3}{c}{Delete-1-MT from Pat $\#1$}    & \multirow{2}{*}{$\widehat{TB}_{d,-1} =r_M\frac{n_{MT,d,-1}}{n_d} + \frac{n_{DLT,d,-1}}{n_d}$ }    \\
  & &  MT & DLT  & \\ \cline{1-5}
    1 &  & \st{1} \bf{0} & 0  &\\ 
    2 & & 1 & 1 & $0.2 \times \frac{1}{3} + \frac{2}{3} = 0.73 $ \\ 
    3 & & 0 & 1   & \\
    \hline
$n_d=3$&  &$n_{MT,d,-1}= \bf{1}$ & $n_{DLT,d,-1}=2$ \\ \hline
  \end{tabular}
  \end{center}
\end{table}
Based on $\widehat{TB}_d$ and $\widehat{TB}_{d,-1}$, PoD-BIN uses a fixed algorithm for Stage I dose-finding summarized in Algorithm~\ref{stageIrules}. The algorithm follows the idea in i3+3 to simplify dose-finding decisions.

\begin{algorithm}[tbh]
\caption{Stage I dose assignment rules $\mathcal{A}(S_d=\emptyset, C_d)$}
\label{stageIrules}
\begin{algorithmic}
\STATE Suppose dose $d$ is currently administered for patients enrolled in the trial.  
\IF{$\widehat{TB}_d$ is below the $EI$}
\STATE  Escalate and  enroll patients at the next higher dose $\mathcal{A}(\emptyset, C_d)=(d + 1)$;

\ELSIF{$\widehat{TB}_d$ is inside the $EI$}

\STATE  Stay and  enroll patients at the current dose $\mathcal{A}(\emptyset, C_d)=d$;

\ELSIF{$\widehat{TB}_d$ is above the $EI$}
\IF{$\widehat{TB}_{d, -1}$ is below the $EI$}
\STATE  Stay and  enroll patients at the current dose $\mathcal{A}(\emptyset, C_d)=d$;
\ELSE
\STATE  De-escalate and  enroll patients at the next lower dose $\mathcal{A}(\emptyset, C_d)=(d-1)$.
\ENDIF
\ENDIF
\end{algorithmic}
\end{algorithm}

To summarize, in Stage I at dose $d$, the unobserved data $S_d=\emptyset$ is an empty set since Stage I only makes a decision when all the patients at dose $d$ complete follow-up and record outcomes. Therefore, the Stage I decision can be described as $A^*_d=\mathcal{A}(S_d=\emptyset, C_d)\in \{d-1,d,d+1\}$, where $\mathcal{A}(\emptyset, C_d)$ represents Algorithm \ref{stageIrules}. This decision not only guides dose-finding in Stage I, but it will also be used in Stage II as well, discussed next.

\paragraph{Stage II } During the first stage, patients are enrolled in cohorts, say, every three patients and newly enrolled patients are assigned to doses following the set of simple dose assignment rules. After each enrollment, check if all four outcomes are observed in the trial or if the number of patients reaches a certain sample size threshold denoted as $n^*$. If the answer is no, the trial continues as Stage I; otherwise, the trial proceeds to Stage II. We recommend using a sample size threshold $n^*$ no less than 12 patients. In Stage II, PoD-BIN switches to full model-based rolling enrollment guided by PoDs. Algorithm~\ref{stageIIrules} provides details. 

The suspension rules in Algorithm~\ref{stageIIrules} are needed to account for the variability of the pending outcomes and protect patient safety. In practice, the values   $\lambda_E$ and $\lambda_D$  should be chosen according to the desired extent of safety and calibrated based on simulations. To ensure safety, we recommend choosing $\lambda_E\geq 0.8$  and setting $\lambda_D\leq0.25$. For example, in the simulation studies (Section \ref{Sim}), we use  $\lambda_E=1$ and  $\lambda_E=0$ , which minimizes the chance of risky decisions while previously enrolled patients have not completed follow-up.  

\begin{algorithm}[H]
\caption{Stage II dose assignment rules}
\label{stageIIrules}
\begin{algorithmic}
\STATE  Suppose dose $d$ is currently administered for patients enrolled in the trial.  Denote $S_d$ the pending data and $C_d$ the observed data at dose $d$.
\IF{there are patients at dose $d$ with pending outcomes}
\STATE 1) Apply the inference based on the Bayesian models in Sections \ref{prob}-\ref{imputation}.
\STATE 2) Compute the PoD in Equation (\ref{eq9}), in which the decision $\mathcal{A}(S_d, C_d)$ is based on Algorithm \ref{stageIrules}. And obtain the optimal decision $A^*_d \in \{d-1, d, d+1\}$. 
\STATE 3) To ensure patient safety, we adopt the following suspension rules \citep{zhou2020pod}. If none of the following suspension rules are invoked, the next cohort of
patients is assigned to the dose indicated by $A^*_d$. Otherwise, suspend the trial until none of the suspension rules are invoked.
  \begin{itemize}
  \item {\bf Suspension rule 1}:  If the current dose $d$ has not been tested before and if the number of patients with pending outcomes exceeds or equals to 3 ($J_d\geq3$), suspend the enrollment.  
  \item {\bf Suspension rule 2}: If $A^*_d=d+1$ (i.e., escalate): the enrollment is suspended if (1) $ Pr(A_d= d+1 | Y_i^{(v_i)},x_i) < \lambda_E$ for some pre-determined threshold $\lambda_E \in [0.33, 1]$ or (2)  if the number of patients who have completed follow-up without DLT is 0.\\ 
 Condition (1) suggests that dose escalation is only allowed if the confidence of escalation is higher than  $\lambda_E$, and a larger $\lambda_E$ represents more conservative escalation decisions. Condition (2) states that escalation is allowed only if at least one patient has finished follow-up with no DLT at the current dose.     
  \item {\bf Suspension rule 3}:  If $A^*_d=d$ (i.e., stay): the enrollment is suspended if $ Pr(A_d= d-1 | Y_i^{(v_i)},x_i) > \lambda_D$  for some pre-determined threshold $\lambda_D \in [0, 0.5]$. This means that stay is allowed only if there is a relatively low chance of de-escalation. A smaller $\lambda_D$ represents more conservative stay decisions.
  \end{itemize}
\ELSE
\STATE  Assign the next cohort of patients according to $\mathcal{A}(\emptyset, C_d)$ in Algorithm \ref{stageIrules}. 
\ENDIF
\end{algorithmic}
\end{algorithm}

\subsection{Safety rules}\label{safety}
In addition to the suspension rules, we include the following safety rules throughout the trial.  Recall that $n_d$ is the number of patients treated at dose $d$, $p_{DLT,d}$ is the DLT probability at dose $d$, and $p^*_{DLT}$ is the target DLT rate. 

\begin{itemize}
\item Safety rule 1 (\textit{early termination}): At any moment in the trial, if $n_1\geq 3$ and $Pr( p_{DLT,1} > p^*_{DLT}|\text{data}) > 0.95$, terminate the trial due to excessive toxicity.
\item Safety rule 2 (\textit{dose exclusion}): At any moment in the trial,  if $n_d \geq 3$ and  $Pr( p_{DLT,d} > p^*_{DLT}|\text{data})>0.95$,  remove dose $d$ and its higher doses from the trial.
\end{itemize}
The posterior probability $Pr(p_{DLT,d}>p^*_{DLT}| \text{data)}$ is calculated using the observed binary DLT data at dose $d$ with a prior distribution $p_{DLT,d}\sim Beta(1,1)$. If Safety rule 1 is triggered in Stages I or II, then the trial is terminated. Safety rule 2 has an exception. Removed doses may be put back to trials if the rule 
is no longer violated after patients with pending data complete follow-up.

\subsection{MTD selection}
\begin{figure}[h!]
\centering
\includegraphics[width=1\textwidth]{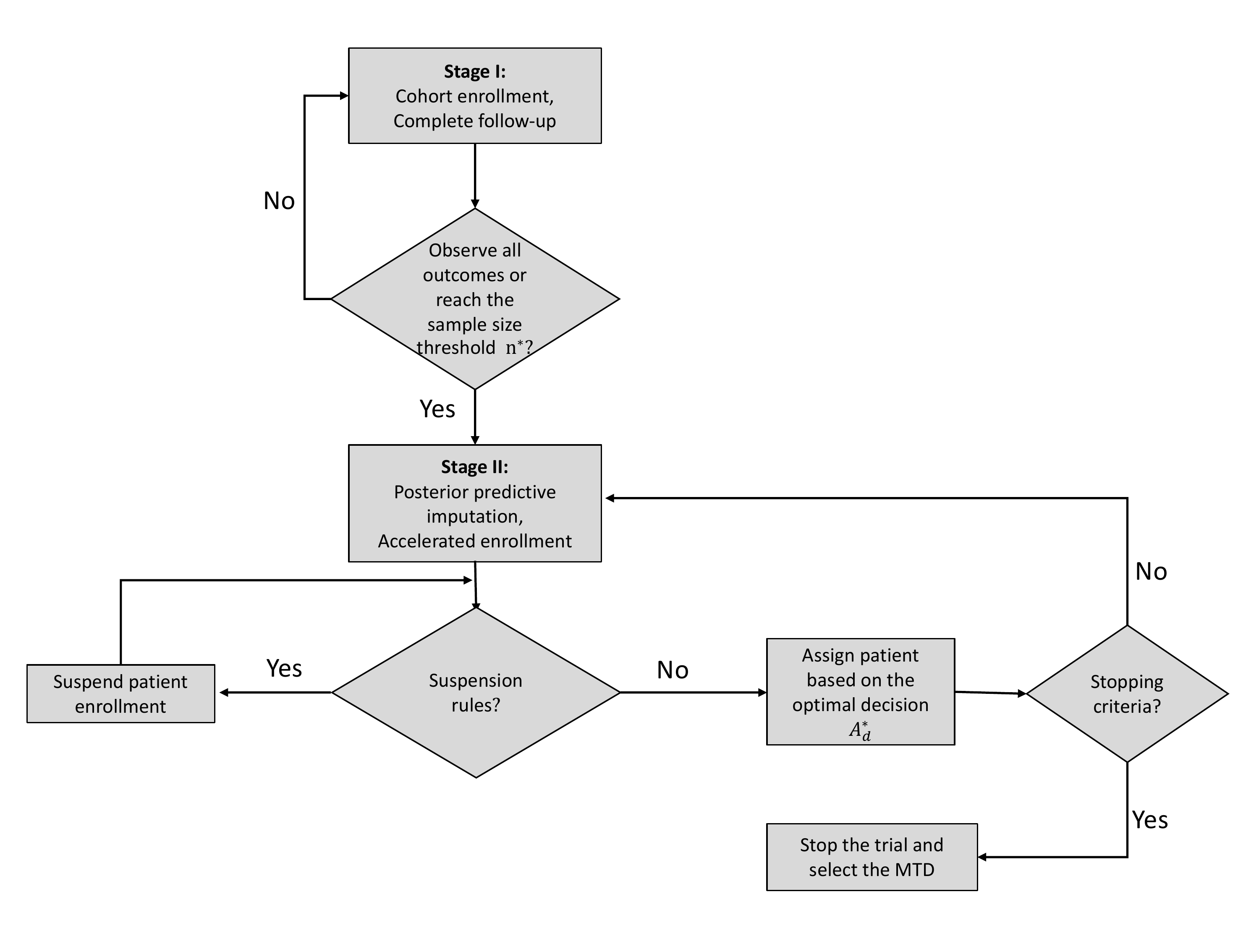}\\
\caption{ A flowchart illustration of the proposed PoD-BIN design. The first cohort is treated at the lowest dose. For subsequent cohorts in Stage I, dose assignments are made based on simple decision rules with complete follow-up data. Once the number of  enrolled  patients reaches sample size threshold $n^*$ or all the ordinal  outcomes  are observed, whichever comes first, Stage II of the design is invoked. This process will be repeated until the stopping criteria are satisfied.}
\label{fig:flowchart}
\end{figure}

The trial is completed if the number of enrolled patients reaches the pre-specified maximum sample size or Safety rule 1 is triggered.  If Safety rule 1 is triggered, no MTD is selected since all doses are considered overly toxic. Otherwise, let $\widetilde{TB}_d=r_M\tilde{p}_{MT,d}+\tilde{p}_{DLT,d}$ where $\tilde{p}_{MT,d}$ and $\tilde{p}_{DLT,d}$ are the posterior means of $p_{MT,d}$ and $p_{DLT,d}$. Let $\{ \widetilde{TB}'_d\}$ be the isotonically transformed $\{ \widetilde{TB}_d\}$ and
PoD-BIN recommends dose $d^\ast$ as the MTD, defined as  
    \begin{equation}
    d^\ast= \argmin_{d\in D,  n_d>0  } |\widetilde{TB}'_d - TTB|.
    \end{equation}
among all doses $d$ for which $n_d>0$.
Figure \ref{fig:flowchart} illustrates the PoD-BIN design as a flowchart.

\vspace{-0.5cm}
\section{ Numerical Studies }\label{sim}
\subsection{Illustration of a single trial in the simulation}

\begin{figure}[h!]
\centering
\includegraphics[scale=0.5]{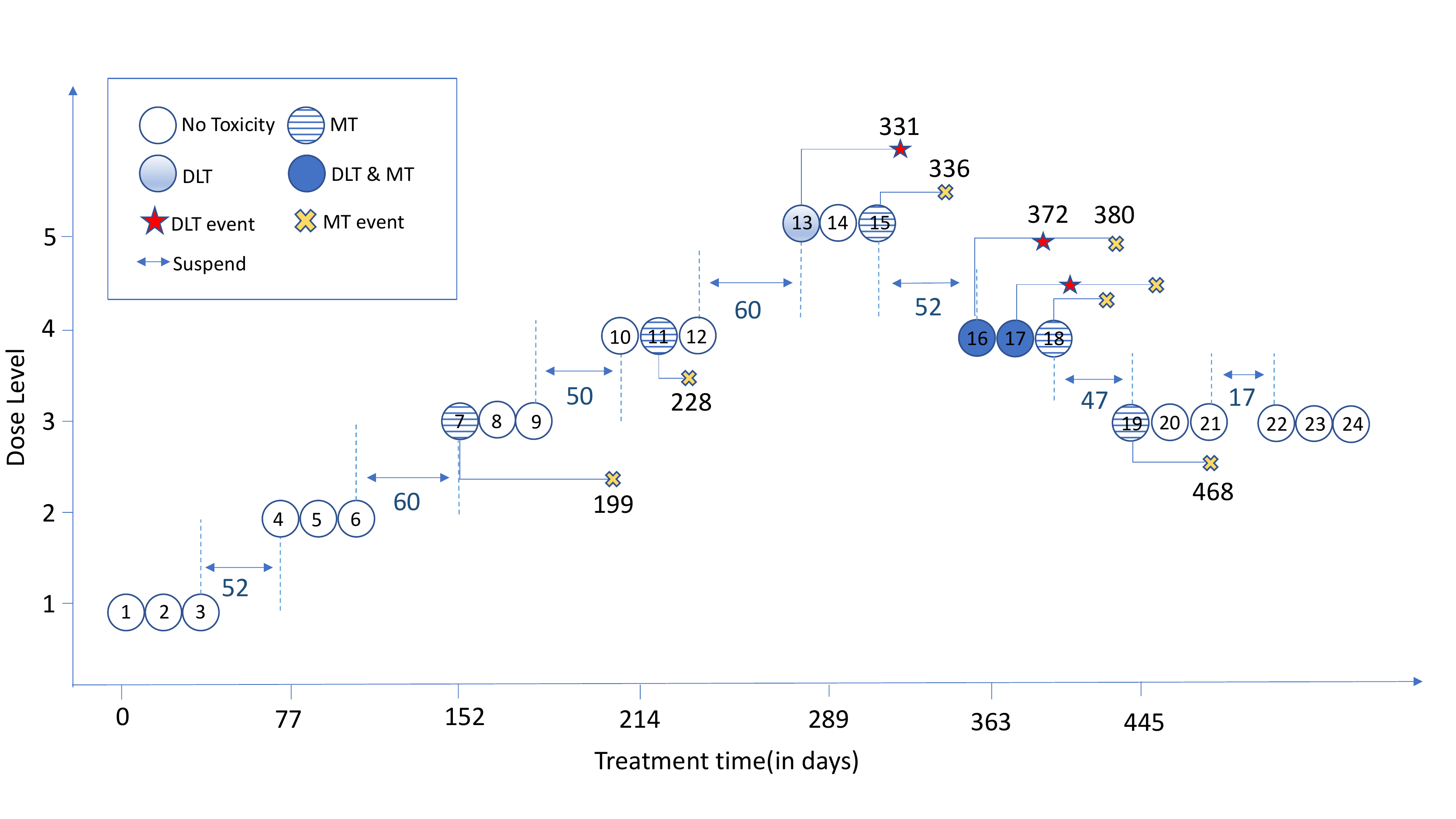}\\
\caption{ A hypothetical dose-finding trial using the PoD-BIN design.}
\label{fig:hypo}
\end{figure}

\noindent  Through a hypothetical dose-finding trial in Figure \ref{fig:hypo},  we illustrate how PoD-BIN is used as a design to guide decision-making. The x-axis is the study time in days, and the y-axis is the dose levels at which patients are treated. Each circle represents a patient.
  Assume    $TTB$ is $0.25$ with $ EI=(0.15, 0.35)$, the assessment window $W=50$ days, 
$r_M=0.15,$  $\lambda_E = 0.95,$   and $\lambda_D = 0.15 $. Patients
are enrolled in cohort sizes of three with a total sample size of
$n=24$. The inter-patient arrival time is generated from an
exponential distribution with a mean of 10 days, i.e., on average, a new patient is enrolled every 10 days. In addition, the sample size
threshold for starting Stage II is set to be $n^*=12$, which is half
of the total sample size. The first 12 patients are enrolled in
cohorts. This means that the patients in the previously enrolled
cohorts must complete the 50-day observation period before the next
cohort of patients can be enrolled. Therefore, no rolling enrollment
is allowed in the first stage.     
Stage II is invoked after the assignment of patient 12 as Stage I
ends. 
Patients $13-15$ are placed at dose level $5$. Patient $13$
experiences DLT, and patient $15$ experiences MT. Hence, dose level 5
is deemed unsafe, and the   next    
cohort (patients $16-18$) is enrolled at dose level 4. 
Later, patients 16, 17, and 18 experience DLT and MT outcomes, which
renders dose level 4 unsafe. Subsequently, the trial de-escalates to
dose level 3,  at which    patients 19 through 24   are enrolled
   according to PoD-BIN.   
 At the end of the trial, dose level 3 is recommended as the MTD by  PoD-BIN,  with 2 MT and 0 DLT events 
 out of 9 patients.  The trial lasts 
 563 days in total.

\subsection{ Simulations}\label{Sim}

We conduct simulation studies to evaluate the operating characteristics of the PoD-BIN design.  The implementation of PoD-BIN requires the specification of a few parameters. We  set 
$TTB = p^*_{DLT}=0.25,$ $ EI=(0.15, 0.35)$, $n^*=12$ and $r_M=0.15$ (so that a DLT event is as severe as more than six MT events). We let  
$\lambda_E = 1$  and $\lambda_D = 0$ for the suspension criteria in
Stage II to minimize the chance of risky decisions.  These two values
are the most conservative choices that   greatly reduce 
the possibility of   making aggressive decisions    
when there are pending patients at the current dose. Less conservative choices will be discussed in the sensitivity studies next. 

We  
compare the proposed PoD-BIN   design    to 
the TITE-CRMMC design \citep{Lee2019}, 
an extension of the TITE-CRM, which allows for the specification of a toxicity target defined based on both DLT and  MT  
constraints.  Importantly,  TITE-CRMMC incorporates the follow-up times of the pending patients using a weighted likelihood approach. For TITE-CRMMC, we apply the recommended model configuration with  
a $N(0, 1.34^2)$ prior for the model parameter, and a skeleton  of
$(0.06, 0.14, 0.25 ,0.38, 0.50)$. Note that in TITE-CRMMC, the
toxicity outcome $Y$ has three levels, $Y = 0$ if no toxicity, $Y=1$
if MTs occur, and $Y=2$ if DLTs occur. There are two toxicity
constraints based on the tail probability of $Y$ under dose $x$, $Pr(Y
\geq 1 | x) = 0.5$ and $Pr(Y = 2 | x) = 0.25$. The next dose level for
TITE-CRMMC is selected based on the marginal posterior of the dose
levels that minimizes the distance to both pre-specified toxicity
constraints, and the MTD recommendation is based on the same
definition. In addition, we add Safety rules 1 \& 2 in Section \ref{safety} to the TITE-CRMMC design for a fair comparison. 
 
Furthermore, we compare the performance of the PoD-BIN design against
a benchmark version of the design, in which Stage I of PoD-BIN is
implemented throughout a trial. In other words, no rolling enrollment
is allowed, and all enrolled patients must be followed for the full
duration with known toxicity outcomes before future cohorts may be
enrolled.    

We consider a total of six different scenarios and simulate trials
investigating five dose levels with a maximum sample size of  
$n=30$    patients and a cohort size of three. The starting dose
level is dose 1. As in   a    typical oncology phase I trial, we
set the length of the assessment window at $W=28$ days. A conditional
uniform model   is    
used to generate the time-to-toxicity in simulations for the proposed
method: we first determine if a patient has a toxic response based on
the true toxicity probability of MT and DLT; and if toxicity is
generated, we generate time-to-toxicity from Uniform$(0,W)$. Moreover,
the arrival time between two consecutive patients follows an
exponential distribution with rate 0.14, meaning on average, one
patient   is enrolled    per week. For each scenario, we simulate 1,000 trials. 
  We also conduct the same simulation with different sample sizes 24
and 36 as a sensitivity study, the results of which are reported in
Web Appendix D.



\subsection{  Operating Characteristics   } 


The simulation results with sample size 30 are summarized in Table \ref{tab:comp1}, which shows the percentage of trials that select each dose as the MTD and the percentage of patients allocated to each dose. 
The true MTD under different designs is bolded. In Scenarios 1 to 3, PoD-BIN outperforms other methods in terms of recommending the correct MTD (PCS). In Scenario 4, the Benchmark method yields slightly better PCS ($80\%$) compared to PoD-BIN($78\%$), with TITE-CRMMC falling behind ($65\%$). In Scenario 5, TITE-CRMMC yields higher PCS compared to PoD-BIN and Benchmark. TITE-CRMMC recommends the correct MTD in $66\%$ of the trials, while PoD-BIN and Benchmark in $61\%$. In Scenario 6, dose levels 4 and 5 are both considered as the true MTD based on PoD-BIN, whereas only dose 5 is considered as the true MTD based on TITE-CRMMC. PoD-BIN and the Benchmark design produce high PCS if dose 5 is considered as the MTD. In terms of allocating patients to the correct dose, PoD-BIN and Benchmark yield similar results and perform well in Scenarios 1,2,5,6. Overall, the performance of PoD-BIN with $\lambda_E = 1, \lambda_D = 0$ is comparable to that of the Benchmark method, in which dose assignments are performed with complete follow-up data.  Across all studied scenarios, the PoD-BIN design results in the best performance in terms of the percentage of trials concluding a dose above the true MTD (POS) and the percentage of patients allocated to doses higher than the MTD (POA), two important safety metrics.

\begin{table}
\scriptsize
\caption{Performance of the PoD-BIN method compared with existing methods. The target toxicity burden is $0.25$ and the $EI$ is $(0.15, 0.35)$.} \label{tab:comp1}
\centering
\tiny
\begin{tabular}{llllllllllll} \hline
                               & \multicolumn{5}{c}{Recommendation percentage}                                           &  & \multicolumn{5}{c}{Allocated dose percentage}                                           \\\hline
\multicolumn{1}{c}{Scenario 1} & 1               & 2               & 3               & 4               & 5               &  & 1               & 2               & 3               & 4               & 5               \\\hline
Pr(MT)                         & \textbf{0.31}   & 0.40            & 0.50            & 0.60            & 0.69            &  & \textbf{0.31}   & 0.40            & 0.50            & 0.60            & 0.69            \\
Pr(DLT)                        & \textbf{0.26}   & 0.38            & 0.50            & 0.62            & 0.74            &  & \textbf{0.26}   & 0.38            & 0.50            & 0.62            & 0.74            \\
TB                             & \textbf{0.31}   & 0.44            & 0.58            & 0.71            & 0.84            &  & \textbf{0.31}   & 0.44            & 0.58            & 0.71            & 0.84            \\\hline
PoD-BIN                        & \textbf{73.0 } & 13.0           & 1.0            & 0.0            & 0.0            &  & \textbf{85.0 } & 13.0           & 2.0            & 0.0            & 0.0            \\
Benchmark                      & \textbf{66.7 } & 16.8           & 1.2            & 0.0            & 0.0            &  & \textbf{82.4 } & 15.2           & 2.3            & 0.2            & 0.0            \\
TITE-CRMMC                     & \textbf{65.0 } & 24.0           & 1.0            & 0.0            & 0.0            &  & \textbf{57.0 } & 32.0           & 9.0            & 2.0            & 0.0            \\\hline
                               &                 &                 &                 &                 &                 &  &                 &                 &                 &                 &                 \\\hline
\multicolumn{1}{c}{Scenario 2} & 1               & 2               & 3               & 4               & 5               &  & 1               & 2               & 3               & 4               & 5               \\\hline
Pr(MT)                         & 0.11            & \textbf{0.26}   & 0.47            & 0.68            & 0.85            &  & 0.11            & \textbf{0.26}   & 0.47            & 0.68            & 0.85            \\
Pr(DLT)                        & 0.07            & \textbf{0.20}   & 0.42            & 0.67            & 0.86            &  & 0.07            & \textbf{0.20}   & 0.42            & 0.67            & 0.86            \\
TB                             & 0.09            & \textbf{0.24}   & 0.49            & 0.77            & 0.98            &  & 0.09            & \textbf{0.24}   & 0.49            & 0.77            & 0.98            \\\hline
PoD-BIN                        & 6.0            & \textbf{86.0 } & 7.0            & 0.0            & 0.0            &  & 37.0           & \textbf{50.0 } & 12.0           & 1.0            & 0.0            \\
Benchmark                      & 3.7            & \textbf{85.9 } & 10.2           & 0.0            & 0.0            &  & 31.4           & \textbf{52.2 } & 15.2           & 1.2            & 0.1            \\
TITE-CRMMC                     & 7.0            & \textbf{69.0 } & 24.0           & 0.0            & 0.0            &  & 15.0           & \textbf{39.0 } & 35.0           & 9.0            & 1.0            \\\hline
                               &                 &                 &                 &                 &                 &  &                 &                 &                 &                 &                 \\\hline
\multicolumn{1}{c}{Scenario 3} & 1               & 2               & 3               & 4               & 5               &  & 1               & 2               & 3               & 4               & 5               \\\hline
Pr(MT)                         & 0.09            & 0.21            & \textbf{0.38}   & 0.56            & 0.70            &  & 0.09            & 0.21            & \textbf{0.38}   & 0.56            & 0.70            \\
Pr(DLT)                        & 0.02            & 0.08            & \textbf{0.21}   & 0.43            & 0.68            &  & 0.02            & 0.08            & \textbf{0.21}   & 0.43            & 0.68            \\
TB                             & 0.03            & 0.11            & \textbf{0.27}   & 0.52            & 0.78            &  & 0.03            & 0.11            & \textbf{0.27}   & 0.52            & 0.78            \\\hline
PoD-BIN                        & 0.0            & 18.0           & \textbf{81.0 } & 1.0            & 0.0            &  & 14.0           & 36.0           & \textbf{39.0 } & 9.0            & 1.0            \\
Benchmark                      & 0.1            & 19.7           & \textbf{79.1 } & 1.1            & 0.0            &  & 13.9           & 33.1           & \textbf{42.2 } & 10.2           & 0.7            \\
TITE-CRMMC                     & 0.0            & 13.0           & \textbf{73.0 } & 14.0           & 0.0            &  & 12.0           & 19.0           & \textbf{42.0 } & 23.0           & 4.0            \\\hline
                               &                 &                 &                 &                 &                 &  &                 &                 &                 &                 &                 \\\hline
                                & \multicolumn{5}{c}{Recommendation percentage}                                           &  & \multicolumn{5}{c}{Allocated dose percentage}                                           \\\hline
\multicolumn{1}{c}{Scenario 4} & 1               & 2               & 3               & 4               & 5               &  & 1               & 2               & 3               & 4               & 5               \\\hline
Pr(MT)                         & 0.11            & 0.21            & \textbf{0.31}   & 0.40            & 0.48            &  & 0.11            & 0.21            & \textbf{0.31}   & 0.40            & 0.48            \\
Pr(DLT)                        & 0.03            & 0.09            & \textbf{0.21}   & 0.37            & 0.57            &  & 0.03            & 0.09            & \textbf{0.21}   & 0.37            & 0.57            \\
TB                             & 0.05            & 0.13            & \textbf{0.25}   & 0.43            & 0.64            &  & 0.05            & 0.13            & \textbf{0.25}   & 0.43            & 0.64            \\
PoD-BIN                        & 1.0            & 18.0           & \textbf{78.0 } & 3.0            & 0.0            &  & 17.0           & 35.0           & \textbf{36.0 } & 11.0           & 1.0            \\
Benchmark                      & 0.3            & 16.0           & \textbf{79.9 } & 3.7            & 0.1            &  & 15.5           & 32.5           & \textbf{38.5 } & 12.1           & 1.5            \\
TITE-CRMMC                     & 0.0            & 12.0           & \textbf{65.0 } & 22.0           & 0.0            &  & 13.0           & 20.0           & \textbf{40.0 } & 22.0           & 4.0            \\
                               &                 &                 &                 &                 &                 &  &                 &                 &                 &                 &                 \\\hline
\multicolumn{1}{c}{Scenario 5} & 1               & 2               & 3               & 4               & 5               &  & 1               & 2               & 3               & 4               & 5               \\\hline
Pr(MT)                         & 0.02            & 0.07            & 0.20            & \textbf{0.37}   & 0.48            &  & 0.02            & 0.07            & 0.20            & \textbf{0.37}   & 0.48            \\
Pr(DLT)                        & 0.00            & 0.00            & 0.03            & \textbf{0.11}   & 0.30            &  & 0.00            & 0.00            & 0.03            & \textbf{0.11}   & 0.30            \\
TB                             & 0.00            & 0.01            & 0.06            & \textbf{0.17}   & 0.38            &  & 0.00            & 0.01            & 0.06            & \textbf{0.17}   & 0.38            \\\hline
PoD-BIN                        & 0.0            & 0.0            & 25.0           & \textbf{61.0 } & 13.0           &  & 10.0           & 11.0           & 19.0           & \textbf{37.0 } & 23.0           \\
Benchmark                      & 0.0            & 1.6            & 23.5           & \textbf{61.4 } & 13.5           &  & 10.1           & 10.6           & 16.9           & \textbf{37.2 } & 25.3           \\
TITE-CRMMC                     & 0.0            & 0.0            & 6.0            & \textbf{66.0 } & 28.0           &  & 11.0           & 10.0           & 13.0           & \textbf{33.0 } & 33.0           \\\hline
                               &                 &                 &                 &                 &                 &  &                 &                 &                 &                 &                 \\\hline
\multicolumn{1}{c}{Scenario 6} & 1               & 2               & 3               & 4               & 5               &  & 1               & 2               & 3               & 4               & 5               \\\hline
Pr(MT)                         & 0.19            & 0.24            & 0.28            & \textbf{0.32}   & \textbf{0.35}   &  & 0.19            & 0.24            & 0.28            & \textbf{0.32}   & \textbf{0.35}   \\
Pr(DLT)                        & 0.04            & 0.06            & 0.08            & \textbf{0.11}   & \textbf{0.15}   &  & 0.04            & 0.06            & 0.08            & \textbf{0.11}   & \textbf{0.15}   \\
TB                             & 0.07            & 0.09            & 0.12            & \textbf{0.16}   & \textbf{0.20}   &  & 0.07            & 0.09            & 0.12            & \textbf{0.16}   & \textbf{0.20}   \\\hline
PoD-BIN                        & 0.0            & 4.0            & 26.0           & \textbf{29.0 } & \textbf{41.0 } &  & 15.0           & 18.0           & 23.0           & \textbf{21.0 } & \textbf{23.0 } \\
Benchmark                      & 0.6            & 6.2            & 22.2           & \textbf{26.3 } & \textbf{44.7 } &  & 15.2           & 18.4           & 20.2           & \textbf{20.7 } & \textbf{25.6 } \\
TITE-CRMMC                     & 0.0            & 3.0            & 23.0           & 40.0           & \textbf{34.0 } &  & 15.0           & 18.0           & 26.0           & 25.0           & \textbf{16.0 } \\\hline
\end{tabular}

\vspace{0.8cm}
 \footnotesize Note: The true MTD under each scenario is bolded.    
\end{table}

\subsection{ trade-off between Safety  and Speed   }\label{trade-off}

  While rolling enrollment can speed up a dose-finding trial, a   
major issue for designs allowing rolling enrollment is the possibility
of making an inconsistent decision. Here, an inconsistent decision
refers to a decision   by the rolling design    that is different
from   its corresponding non-rolling design. In other words, an
inconsistent decision by the rolling design is different from the
decision that 
would have been made if all patients had been followed for the full
duration with complete toxicity outcomes.  For PoD-TPI and other
rolling designs,  there can be    six types of inconsistent decisions
  denoted as AB, where A is the decision that should have been made
if complete toxicity data were available, and B is the decision made by
the rolling design allowing pending toxicity outcomes. The six types
are:    (1) should de-escalate but stayed (DS), (2) should
de-escalate but escalated (DE), (3) should stay but escalated (SE),
(4) should stay but de-escalated (SD), (5) should escalate but
de-escalated (ED), and (6) should escalate but stayed (ES).   For
example, SE refers to an inconsistent decision of E, escalate, made by
the rolling design while the corresponding non-rolling design based on
complete data would make the decision S, stay at the current
dose. Apparently, SE is a risky decision since the rolling design
wrongly exposes patients to a high dose.     
  Three inconsistent decisions   DS, DE and ES are considered as
{\bf risky decisions}, as they can expose patients to overly toxic
doses, and SD, ED and ES are conservative decisions, which could
allocate patients to sub-therapeutic doses.   In general, the more
liberal a rolling design allows incomplete data to inform doses for
future patients, the higher risk the design has in putting patients at
overly toxic doses. To balance the trade-off of speed and safety,
PoD-BIN uses    the suspension thresholds $(\lambda_E,\lambda_D )$ to
control the chance of making these inconsistent decisions.    

We   report simulation results of PoD-BIN    
with different combinations of the suspension threshold values
$(\lambda_E,\lambda_D)$. For each scenario and suspension threshold,
we simulate 1,000 trials using PoD-BIN. 
Table \ref{Sen1} 
reports the mean and standard deviation of different metrics over the
six  scenarios with 1,000   simulated    trials for each scenario. 
The scenario-specific operating characteristics are reported in  Web
Appendix E. We calculate seven different metrics to summarize the  
operating characteristics of the PoD-BIN with different thresholds,    shown in Panel A of Table \ref{Sen1}. 
In addition, we report the percentage of inconsistent decisions to
evaluate the reliability and safety of PoD-BIN   in Panel B.   

Table \ref{Sen1} Panel A shows comparable PCS and POS of PoD-BIN to
those of Benchmark since the selection of MTD is always based on
complete data. The POA, the average percentage of patients who experience
MT (POMT), and the average percentage of patients who experience DLT
(PODLT) of PoD-BIN are slightly better compared to Benchmark. Panel B
shows the effect of $\lambda_E$ and $\lambda_D$ on limiting aggressive
decisions. The closer to 1   is    for $\lambda_E$ and to 0 for
$\lambda_D$, the safer the PoD-BIN but also the longer the trial.
Importantly, the trial duration can be greatly reduced by using the
PoD-BIN method   in all cases.    On average, the trial duration is shortened by at least 50 days using PoD-BIN compared to Benchmark. 
An important feature of PoD-BIN is the flexibility in calibrating the
values of $\lambda_E$ and $\lambda_D$ to achieve the desired trade-off
between speed and safety. As noted in \cite{zhou2020pod}, the
threshold $\lambda_E$ controls the inconsistent of DE and SE
decisions, and $\lambda_D$  controls the rate of SD decisions. They
both affect the speed of the trial through suspension rules. With
$\lambda_E=1, \lambda_D=0$, PoD-BIN achieves the lowest POA, POMT and
PODLT, and the numbers of risky decisions (DS, DE, SE) are
minimized. On the contrary, by setting $\lambda_E=0.85,
\lambda_D=0.25$, the design produces   higher risks of making
aggressive    decisions. However, the average trial duration is further shortened with less strict threshold values, which is a result of less frequent and shorter periods of suspension during a trial. The trade-off between trial speed and safety can be carefully adjusted by choosing an appropriate threshold for suspension in practice, a distinctive and important feature of PoD-BIN.\\


\begin{table}[hbtp]
\caption{ Performance of PoD-BIN with different choices of $\lambda_E, \lambda_D$.   $\mbox{Mean}_{\mbox{sd}}$ values are shown in all the entries.    }\label{Sen1}
\centering
\small
\subcaption*{Panel A: PCS: the average percentage of correct selection of the true MTD; PCA: the average percentage of patients correctly allocated to the true MTD; POS: the average percentage of selecting a dose above the true MTD; POA: the average percentage of patients allocated to a dose above the true MTD; POMT: the average percentage of patients experiencing moderate toxicities; PODLT: the average percentage of patients experiencing dose-limiting toxicities. The unit of PCS, PCA, POA, POMT, and PODLT is $\% $.}
\begin{tabular}{lcccccc}
\hline
     & PCS   & PCA   & POA   & POS  & POMT  & PODLT  \\\hline
     Benchmark & $74.00_{9.20}$ &	$49.77_{16.89}$ &	$13.96_{8,39}$ &	$7.77_{7.27}$&	$28.39_{2.51}$&	$16.68_{5.64}$ \\
$ \lambda_E = 1,\lambda_D = 0$  & $\bf{74.90_{8.95}}$  & $48.62_{18.74}$ & $\bf{12.18_{7.34}}$ & $\bf{6.37_{6.22}}$ & $27.75_{3.15}$ & $16.44_{5.77}$ \\
$ \lambda_E = 0.95,\lambda_D = 0$  & $74.10_{10.27} $& $48.93_{18.30}$ & $12.54_{7.41}$ & $6.78_{6.51}$ & $\bf{27.65_{2.78}}$& $\bf{16.38_{5.73}}$ \\
 $\lambda_E = 1,\lambda_D = 0 .15$ & $74.02_{9.09}$ & $49.29_{18.26}$ & $12.55_{7.85}$ & $6.67_{6.25}$ & $27.66_{2.62}$ & $16.62_{5.86}$ \\
 $\lambda_E = 0.95,\lambda_D = 0 .15$ & $74.60_{9.13}$ & $49.02_{18.02}$ & $12.78_{7.63}$ & $6.47_{5.95}$ & $27.84_{2.90}$  & $16.60_{5.82}$  \\
 $\lambda_E = 0.85,\lambda_D = 0 .25$ & $73.53_{9.34}$ & $\bf{49.80_{17.46}} $ & $13.75_{8.22}$ & $7.38_{6.97} $& $28.11_{2.67}$ & $16.92_{5.86}$\\\hline
\end{tabular}
\vspace{0.9cm}
\subcaption*{Panel B: Frequencies of inconsistent decisions and duration. The unit of inconsistent decisions (DS, DE, SE, SD, ED, ES) is $1/100$, and the unit of duration (Dur) is day. }

\begin{tabular}{lccccccc}
\hline
     & DS   & DE   & SE   & SD   & ED   & ES  & Dur \\\hline
     Benchmark  & -  & -   & -  & -  & - & - &  $476_{20}$ \\
 $\lambda_E = 1,\lambda_D = 0 $ & $\bf{0.37_{0.57}}$ & $\bf{0.01_{0.01}}$& $\bf{0.01_{0.01}}$ & $3.87_{2.12}$ & $0.27_{0.29}$ & $\bf{1.60_{0.76}}$  & $423_{24}$ \\
 $\lambda_E = 0.95,\lambda_D = 0 $ & $0.46_{0.79}$ & $0.06_{0.04}$ & $0.81_{0.41}$ & $3.97_{2.18}$ & $0.32_{0.23}$ & $1.28_{0.58}$ & $411_{25}$ \\
 $\lambda_E = 1,\lambda_D = 0 .15$ & $2.38_{0.53}$ & $\bf{0.01_{0.01}}$ & $\bf{0.01_{0.01}}$ & $3.05_{1.60}$ & $0.18_{0.15}$ & $3.90_{1.84}$ & $392_{28}$ \\
$ \lambda_E = 0.95,\lambda_D = 0 .15$ & $2.43_{0.70}$ & $0.07_{0.04}$ & $0.87_{0.44}$ & $2.93_{1.55}$ & $0.20_{0.15}$ & $3.44_{1.69}$ & $383_{26}$ \\
$ \lambda_E = 0.85,\lambda_D = 0 .25$ & $3.18_{0.70}$ & $0.21_{0.19}$ & $1.82_{0.66}$ & $\bf{2.82_{1.56}}$ & $\bf{0.17_{0.15}}$ & $2.67_{1.30}$ & $\bf{376_{20}} $\\\hline
\end{tabular}
\vspace{0.8cm}
 \footnotesize 
Note: Bold values indicate the optimal performance among all cases.
\end{table}

\subsection{Sensitivity Analysis}
Additionally, we conduct sensitivity analysis to demonstrate the
performance of PoD-BIN with different generative models to simulate
patients' time-to-toxicity profiles and trial accrual rates. First,
assuming patient accrual time follows $Exp(1/7)$ and assessment window
length $W=28$ days, we simulate time to MT/DLT for a patient from a
Weibull distribution with shape and scale parameters uniquely
identified based on the following four settings.   See    Web Appendix G for details.\\
\indent  $\bullet$ Setting 1: $80\%$ of DLTs occur in the first half of the assessment window $W=28$ days, and $80\%$ of MTs occur in the first half of the assessment window days;\\
 \indent $\bullet$ Setting 2: $20\%$ of DLTs occur in the first half of the assessment window $W=28$ days, and $20\%$ of MTs occur in the first half of the assessment window days;\\
\indent  $\bullet$ Setting 3: $80\%$ of DLTs occur in the first half of the assessment window $W=28$ days, and $20\%$ of MTs occur in the first half of the assessment window days;\\
\indent  $\bullet$  Setting 4: $20\%$ of DLTs occur in the first half of the assessment window $W=28$ days, and $80\%$ of MTs occur in the first half of the assessment window days. \\
For each scenario and time-to-toxicity setting, we simulate 1,000 trials. The sensitivity results with four time-to-toxicity settings are summarized in Table \ref{Sen2}. Setting 1 suggests that both DLT and MT occur early during the follow-up period, and it achieves the shortest trial duration with the lowest POMT and PODLT. On the other hand, setting 2 indicates that both DLT and MT occur late during the follow-up period, and it achieves the longest trial duration with the highest POMT and PODLT.  In terms of inconsistent decisions, both settings 2 and 4 with late DLT occurrence tend to have higher frequencies of risky decisions DS, DE and SE comparing to the other two settings with early DLT occurrence.

\begin{table}[H]
\caption{ Performance of PoD-BIN under different time-to-toxicity
  profiles.   $\mbox{Mean}_{\mbox{sd}}$ alues are shown in all the entries.    
}\label{Sen2}
\centering
\small
\subcaption*{Panel A: PCS: the average percentage of correct selection of the true MTD; PCA: the average percentage of patients correctly allocated to the true MTD; POS: the average percentage of selecting a dose above the true MTD; POA: the average percentage of patients allocated to a dose above the true MTD; POMT: the average percentage of patients experiencing moderate toxicities; PODLT: the average percentage of patients experiencing dose-limiting toxicities. The unit of PCS, PCA, POA, POMT, and PODLT is $\% $.}

\begin{tabular}{lllllll}\hline
                   & PCS   & PCA   & POA   & POS  & POMT  & PODLT \\\hline
Setting 1         & $74.42_{8.23}$	&$48.46_{18.74}$&	$12.22_{7.17}$&	$6.80_{6.47}$ &	$\bf{27.40_{2.93}}$&	$\bf{16.21_{5.82}}$	\\
Setting 2          & $\bf{75.32_{10.14}}$	&$48.74_{18.22}$	&$12.39_{7.35}$&	$\bf{6.38_{6.40}}$&$27.96_{3.32}$&$16.57_{5.94}$\\
Setting 3  & $74.73_{10.95}$&	$\bf{49.32_{19.19}}$&	$\bf{12.18_{7.60}}$&	$6.68_{6.11}$	&$27.69_{2.84}$&	$16.43_{5.87}$	\\
Setting 4  &$ 74.24_{8.82}$	&$48.70_{18.68}$&	$12.49_{7.62}$&	$6.77_{6.54}$	&$27.74_{2.75}$&	$16.44_{5.76}$	\\\hline
\end{tabular}
\vspace{0.9cm}
\subcaption*{Panel B: Frequencies of inconsistent decisions and duration. The unit of inconsistent decisions (DS, DE, SE, SD, ED, ES) is $1/100$, and the unit of duration (Dur) is day. }

\begin{tabular}{llllllll}\hline
&DS            & DE            & SE            & SD            & ED            & ES      & Dur \\\hline

Setting 1         & $\bf{0.26_{0.53}}$&	$0.01_{0.01}$	&$\bf{0.01_{0.02}}$&	$4.56_{2.68}$&	$0.38_{0.34}$&$1.99_{0.95}$          &$\bf{415_{27}}$\\
Setting 2          & $0.46_{0.83}$	&$0.01_{0.01}$&	$0.02_{0.01}$	&$\bf{3.43_{1.88}}$&	$0.21_{0.19}$	&$\bf{1.29_{0.68}}$ &$429_{21}$\\
Setting 3  &    $0.27_{0.56}$&	$\bf{0.00_{0.01}}$&	$\bf{0.01_{0.01}}$&$3.95_{2.30}$&$\bf{0.20_{0.15}}$&$1.41_{0.81}$  &$423_{26}$\\
Setting 4  & $0.55_{0.73}$&	$0.02_{0.02}$&$0.05_{0.03}$	&$3.97_{2.34}$&	$0.43_{0.44}$&$1.67_{0.83}$  &$422_{24}$\\\hline
\end{tabular}
\vspace{0.8cm}
 \footnotesize 
Note: Bold values indicate the optimal performance among all settings.
\end{table}

\vspace{-0.5cm}
\section{Discussion}\label{dis}

In this article, we have proposed the two-stage PoD-BIN design to incorporate time-to-toxicity data of multiple toxicities to speed-up phase I trials. The proposed design defines the concept of toxicity burden to summarize multiple toxicity outcomes information simultaneously and adopts a decision algorithm similar to the i3+3 design \citep{Liu2020} in Stage I. Furthermore, the proposed PoD-BIN design evaluates the posterior predictive probabilities of dose escalation decisions by using a latent probit model in Stage II. Flexible suspension rules based on the risk of different dosing decisions are added to further ensure trial safety and control the chance of making inconsistent decisions. The thresholds for suspension can be adjusted to balance the trade-off between speed and safety, as demonstrated in Section~\ref{trade-off}. 

We considered a two-stage design, in which Stage I acquires complete follow-up data from patients, and Stage II allows rolling enrollment based on model inference. There are two criteria to invoke Stage II: 1) once the sample size reaches a pre-specified threshold $n^*$ , or 2) once all 4 toxicity outcomes are observed. We recommend $n^*$ no less than 12 patients. In practice, the sample size threshold $n^*$ can be determined through sensitivity analysis to evaluate its effect on the design performance.  
Additionally, the severity weight that quantifies the relative severities of MT and DLT must be elicited from the physicians planning the trial. These decisions are inherently subjective and require close collaboration between the statisticians and the clinical team.  Multiple physicians are recommended to quantify this severity weight, and additional sensitivity analysis can be implemented to evaluating its impact on the operating characteristics of the design.
 
PoD-BIN delivers sound operating characteristics that are comparable to existing designs. Comparing to the TITE-CRMMC design, the proposed design tends to reduce the probability of selecting a dose having excessive toxicities and the number of patients allocated to dose above the MTD. Using the information on moderate toxicities to guide the estimation of MTD seems to result in conservative dose-finding decisions in comparison with standard phase I designs.

In a real trial, it is desirable to account for both times to toxicities as well as the number of occurrences of different toxicities. Future research is needed to extend the model to handle count data.


\newpage

\bibliographystyle{apalike}
\bibliography{reference}

\end{document}